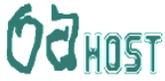

# Magnetic Force Microscopy Characterization of Superparamagnetic Iron Oxide Nanoparticles (SPIONs)

Gustavo Cordova[1], Simon Attwood[2], Ravi Gaikwad[2], Frank Gu[3,4], Zoya Leonenko[1,2,4,]

[1]Department of Biology, University of Waterloo, Waterloo ON
[2]Department of Physics and Astronomy, University of Waterloo, Waterloo ON
[3]Department of Chemical Engineering, University of Waterloo, Waterloo ON
[4]Waterloo Institute for nanotechnology, University of Waterloo, Waterloo ON

✉ Corresponding author: E-mail: zleonenk@uwaterloo.ca, Tel.: 519-888-4567 x38273

## Abstract

Superparamagnetic iron oxide nanoparticles (SPIONs), due to their controllable sizes, relatively long *in vivo* half-life and limited agglomeration, are ideal for biomedical applications such as magnetic labeling, hyperthermia cancer treatment, targeted drug delivery and for magnetic resonance imaging (MRI) as contrast enhancement agents. In order to understand how SPIONs interact with cells and cellular membranes it would be of interest to characterize individual SPIONs at the nanoscale in physiologically relevant conditions without labeling them. We demonstrate that Magnetic Force Microscopy (MFM) can be used to image SPIONs in air as well as in liquid. The magnetic properties of bare and $SiO_2$ coated SPIONs are compared using MFM. We report that surface modification using (3-mercaptopropyl)-trimethoxysilane significantly improves adsorption and distribution of SPIONs on gold surfaces. To obtain proof of principle that SPIONS can be imaged with MFM inside the cell we imaged SPIONs buried in thin polymer films (polystyrene (PS) and poly methyl-methacrylate (PMMA)). This opens the possibility of visualizing SPIONs inside the cell without any labeling or modifications and present MFM as a potential magnetic analogue for fluorescence microscopy. The results of these studies may have a valuable impact for characterization and further development of biomedical applications of SPIONs and other magnetic nanoparticles.

## Introduction

Biocompatible superparamagnetic iron oxide nanoparticles (SPIONs) have been widely used for biomedical applications such as tissue specific release of therapeutic agents, magnetic hyperthermia treatment for cancer patients, a wide range of cell separation techniques as well as contrast agents in MRI imaging. Inorganic coatings, such as aluminum, cadmium, gold and silica can be used to electrostatically stabilize magnetic nanoparticles in a colloid [1]. These inorganic materials are most commonly used for post-synthesis modification of SPIONs which will adopt the core-shell nanoparticle structure. Recently, silica has received a great deal of attention for this purpose.

Due to its biocompatibility, low cost, and allowance for covalent stabilization over a broad range of pHs, silica (ie. silicon dioxide) has become an ideal choice for post-synthesis modification of SPIONs to be used for biomedical applications [2, 3]. Silica is highly suitable for preserving the intrinsic magnetic properties of SPIONs by helping to prevent oxidation and aggregation of the SPION's magnetite core [4]. Although, in 2008 a study by Bumb *et al.* used SQUID magnetic analysis and showed that under low applied fields, higher magnetization values were observed for the silica SPION sample as compared to uncoated SPIONs, suggesting that silica separating the small particles may be leading to weak ferromagnetic ordering in the relatively large batches of nanoparticles that are required for SQUID analysis [4].

Today, SPIONs are popularly used in a large variety of therapeutic and diagnostic biomedical applications, both *in vitro* and *in vivo*. Most often SPIONs are used as magnetic resonance imaging (MRI) contrast enhancement agents. They are intravenously infused into the body to detect and characterize small lesions, tumours in organs, or to visualize the digestive tract [5]. Due to their high magnetization, SPIONs cause a critical decrease in the relaxation rate of water protons, and therefore are efficient MRI contrast agents. The enhanced contrast allows MRI to differentiate between different organs in the body as well as benign and malignant tissues [6]. Iron oxide nanoparticles are used most commonly for this purpose due to their low toxicity, chemical stability and biocompatibility.

Although MRI is a powerful technique, its resolution is in the range of millimeters to micrometers and it does not give information about position of SPIONs at a single cell level. In order to develop better SPION-based contrast agents it is important to understand how SPIONs interact with the cell and cellular membrane at the nanoscale without labeling them. Currently, one of the most common methods for intracellular imaging of magnetic nanoparticles is fluorescence microscopy. A disadvantage of this technique is that nanoparticles must first be labeled or modified with fluorescent probes in order for the particles to be visualized, which may affect their interaction with the cellular membrane. Furthermore, the maximum resolution of this technique is limited to 300-500nm - half the wavelength of the light being used [7]. Relative to fluorescence microscopy, two-photon microscopy (TPM) offers improved resolution and has also been used to study cellular interactions with magnetic nanoparticles but still requires the particles to be labeled with a two-photon fluorescent dye [8]. Due to the relatively poor resolution and reliability of these techniques, a label-free *in vitro* detection method for magnetic nanoparticles, SPIONs especially, is of great interest. Magnetic force microscopy (MFM), because of its ability to localize and characterize magnetic nanoparticles at the nanoscale without labeling, offers great potential for this purpose.

MFM has the capability to detect nanoscale magnetic domains as well as simultaneously obtain both atomic force microscopy phase and topography images. This technique has received limited attention as a potential tool for characterization of SPIONs, specifically in physiologically relevant conditions, and has only ever been used in liquid to image computer hard disks [9]. Most studies that have used MFM to characterize SPIONs, or other magnetic nanoparticles, have done so under ambient conditions (in air) [10, 11, 12, 13]. SPIONs, however, when applied in biomedicine, typically carry out their function in physiological conditions. Therefore, characterization of these particles should be undertaken in conditions similar and relevant to the physiological environment, i.e. in liquid. The potential of MFM is largely unexplored in this regard.

In this study, we evaluate the applicability of MFM in air, as well as in liquid, to characterize bare and $SiO_2$ coated SPIONs on mica. The magnetic properties of individual bare and $SiO_2$ coated SPIONs are compared using MFM. For the first time we demonstrate that MFM imaging of SPIONs can be done in liquid. To mimic SPIONs buried inside the cell we imaged them inside thick polymer film (polystyrene (PS) and poly-methyl methacrylate (PMMA)). This will provide a platform for cellular studies on SPIONs without any labeling.

## Experimental Section
### SPION Synthesis
**Materials**

Iron (II) chloride (99%, Sigma), iron (III) chloride

(99%, Sigma), tetramethylammonium hydroxide (TMAOH) (25% solution, Sigma), tetraethyl orthosilicate (TEOS) (99%, Sigma), ammonium hydroxide (28% solution, Sigma), and ethanol (99%, ACS) were purchased and used without further purification. (3-Mercaptopropyl) trimethoxysilane (MPTS) was purchased from Sigma (95%).

**SPION Synthesis**

Bare and $SiO_2$ coated SPIONs were synthesized by a co-precipitation method as described previously [14]. Briefly, 2.5 mL of a mixed iron solution in deionized water (2 mol/L $FeCl_2$ and 1 mol/L $FeCl_3$) was added to a 0.7 mol/L tetramethylammonium hydroxide (TMAOH) solution under vigorous stirring, and the reaction was allowed to proceed open to the air at room temperature for 30 minutes while stirring. After 30 minutes, the black particles were separated from solution over a neodymium magnet, and washed at least thrice with an equivalent volume of pH 12 TMAOH solution (so as to maintain the equivalent particle concentration as immediately after the reaction) until the particles were no longer magnetically separable. This colloidal suspension was sonicated for 10 minutes (Branson Digital Sonifier 450, USA), and then 20 mL of the sonicated fluid was mixed with 20 mL pH 12 TMAOH and 160 mL ethanol. 7 mL tetraethylorthosilicate (TEOS) was then added to this suspension while stirring, and allowed to react at room temperature while stirring for approximately 18 hours. The $SiO_2$ coated SPIONs were then magnetically recovered from solution, and washed thrice with ethanol and thrice with deionized water by magnetic decantation, and sonicated in deionized water for 10 minutes before further use.

## Deposition of SPIONs on mica substrate for imaging

A SPION dilution of ~5.5 mg/mL was prepared using deionized water. A concentration of ~5.5 mg/mL of SPIONs was used because it was observed to provide a uniform distribution of SPIONs with a relatively small size distribution where individual particles could be observed. The SPION dilution was sonicated for 25 minutes. After this 10 μL of the SPION dilution was deposited on freshly cleaved mica (v-4 grade, SPI Supplies, PA, USA), covered with a petri dish and left to air dry for approximately six minutes. The sample was gently rinsed with four or five drops of deionized water and then immediately dried with a steady stream of nitrogen for 3.5 minutes, placed in a sealed petri dish with nitrogen and left in a dessicator overnight.

For polymer coated samples, a three percent (3%) solution of PMMA or 0.4% solution of polystyrene was used to spin coat the SPION samples with a coating of ~30 nm. Both PMMA and polystyrene dilutions were made in toluene. 40 μL aliquots of the respective polymer solutions were pipetted onto the already deposited SPION samples. A spin motor with an applied voltage of 1V for 15 seconds was used in order to spin coat both the PMMA and polystyrene.

**Substrate (mica) modification via 3-MPTS**

Circular mica substrates were freshly cleaved and stored in nitrogen. The mica substrates were then sputtered with a 2.5 nm layer of titanium using an electron beam evaporator. Subsequently, a 50 nm layer of gold was deposited onto the titanium layer without interruption of the vacuum ensuring that titanium dioxide did not form on the surface of the substrate. After sputtering, mica substrates were stored in nitrogen until needed. Sputtered gold wafers were immersed in a 40 mM solution of 3-MPTS in methanol for three hours. Substrates were then thoroughly rinsed with methanol and Millipore water. After immersion in methanol, substrates were then placed into an aqueous 0.01 M NaOH solution for one hour. Substrates were then immersed in a solution of $SiO_2$ SPIONs (~6.9 mg/mL) for one hour. To finish, substrates were washed with Millipore water and dried with nitrogen gas. Samples were imaged immediately.

**MFM imaging**

All samples were imaged using a Nanowizard II atomic force microscope (JPK Instruments, Germany). MikroMasch NSC-18, cobalt/chromium coated magnetic cantilevers (MIKROMASCH CA, USA) were used to image the SPIONs. An external perpendicular magnetic field was applied to the sample during imaging in order to ensure magnetization of the SPIONs and improve the contrast of MFM phase images by placing a small permanent magnet directly underneath the mica during imaging [10]. All samples were imaged using AFM intermittent contact mode with MFM hover mode (or lift mode) at various distances from the substrate.

## Quantitative Analysis of Topography and MFM Images

All particle analysis was conducted using customized scripts written in Matlab. Raw images were flattened by subtracting a best fit plane (plane flatten). The median of each row was then found, and subtracted from each pixel in that row (line-by-line flatten). Next a course height (or magnetic signal if assessing the MFM image) threshold was used to highlight the features not associated with the flat substrate. The raw data was then re-loaded and a plane-flatten and line-by-line flatten were employed whilst excluding the previously highlighted regions. This procedure allowed the images to be correctly flattened which is extremely important for particle analysis. A sum of two Gaussians was then fitted to the histogram of height (or magnetic signal) data. For the topography images the position of the first peak represents the mica surface whereas for the MFM images this represents the background magnetic signal. Often a second height (or MFM) threshold was employed so as to carefully highlight the pixels associated with the particles. The maximum height for each particle was then measured relative to the mica surface from the topography image. The maximum magnetic signal shift was measured relative to the background signal from the MFM image. Thus our Matlab routine allows both the maximum height and corresponding magnetic signal to be assessed for each particle.

## Results and Discussion
### Bare vs. SiO$_2$ coated SPIONs

In order to compare the magnetic properties of bare and SiO$_2$ SPIONs at the nanoscale, MFM imaging was done on samples of each type of SPION deposited on mica as described in section 2.2. Panels A - D in Figure 1 show the AFM and MFM phase images of the bare and SiO$_2$ coated SPIONs. AFM topography images (panels A and C) were analyzed to obtain the size distributions for both types of SPIONs. The distribution of particle size (diameter) is shown in Figure 1E and 1F and was determined using height distribution analysis. Both bare and silica coated SPIONs have non-symmetric right skewed size distributions with the SiO$_2$ SPIONs having a broader distribution and a higher mean diameter than the bare SPIONs.

The mean, median and mode diameter measured for bare and SiO$_2$ coated SPIONs were 5.1 +/− 0.1 nm (standard error), 4.0 nm, 1.2 nm and 33 +/− 1 nm (standard error), 13 nm, 4 nm respectively. Although MFM detection of SPIONs has been reported, there is some doubt whether individual SPIONs can actually be distinguished by MFM because the magnetic field from SPIONs is proportional to the diameter of the particle and thus very small. In 2009, silica nanoparticles, with and without the presence of a magnetic core were compared using MFM [15]. When the magnetic core was absent, no MFM contrast was observed suggesting that only magnetic structures will cause measurable phase contrast. In Figure 1, MFM contrast is observed for SPIONs as small as ~3 nm. As a result, we can be confident that these are true MFM signals from the SPIONs being studied.

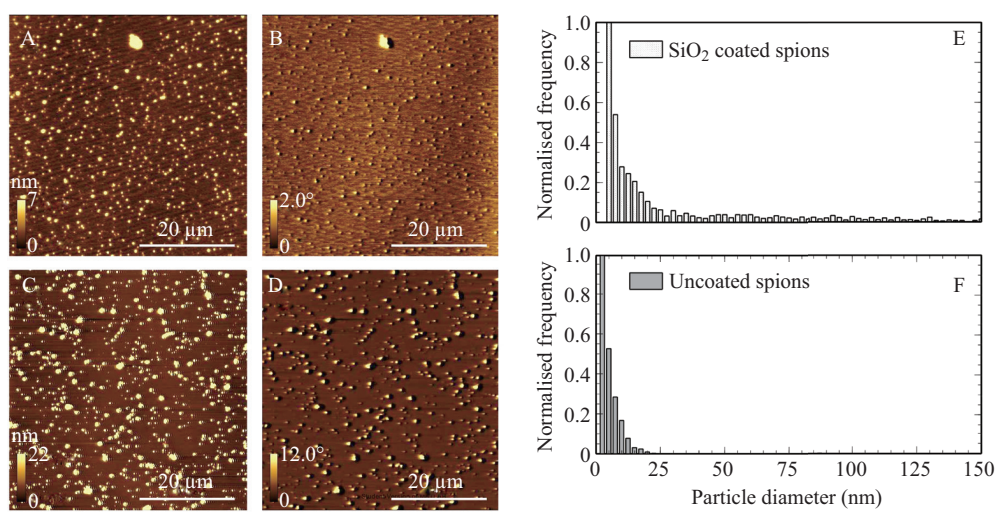

**Fig. 1** Characterization of bare vs. SiO$_2$ coated SPIONs. A-D. Tapping-mode AFM topography images (column 1) and MFM phase images (colum 2) of bare and SiO$_2$ coated SPIONS (images A/B and C/D respectively). Both topography and MFM phase images obtained using a magnetic AFM probe in the presence of an externally applied perpendicular magnetic field in ambient conditions. MFM phase images obtained in lift mode at a scan height of 50 nm. Colour-scale and scale bars for both topography and phase images are shown in the bottom of each panel. E/F) Size distributions for both bare (E) and SiO$_2$ coated (F) SPIONs.

The contrast observed in the MFM images is caused by the interaction between MFM probe and the magnetic field from the SPIONS. These interactions cause a shift in the phase of the oscillating probe [16, 17]. The MFM images for both types of SPIONs shown in Figure 1, panels B and D demonstrate an effect called dipolar contrast, with half of the phase contrast being dark, and half being light for each individual magnetic structure. This dipolar contrast for magnetic nanoparticles is typically found only when external magnetic fields are applied perpendicularly to the measurement direction as is the case for this study [10, 15, 18, 19].

In order to directly compare the magnetic properties of individual bare and $SiO_2$ coated SPIONs, the MFM phase-shift for both bare and $SiO_2$ SPION, gathered from the MFM phase images (Figure 1, panels B and D), was plotted as a function of particle size (diameter) in Figure 2. A positive linear trend is observed for both types of SPIONs. This data suggests that the magnetic moment is proportional to the diameter (and therefore the volume) of the SPION. Computer simulations of MFM on SPIONs have also demonstrated that the phaseshift detected in MFM depends very strongly on the particle diameter [18]. A 2008 study by Bumb *et al.,* noted that under low applied fields, higher magnetization values were observed for silica-coated SPION samples when compared to uncoated SPIONs; suggesting that silica separating the small particles may be leading to weak ferromagnetic ordering in the relatively large batches of nanoparticles that are required for SQUID analysis [4]. We found bare and $SiO_2$ coated SPIONs behave identically when analyzed with MFM, demonstrating that the $SiO_2$ coating has no effect on the magnetic properties of the SPIONs – contrary to large batch analysis using SQUID. A similar result was also observed by Neves *et al.* in 2010 which found that the response of MFM to magnetic nanoparticles is not affected by the presence of a silica coating [19].

## MFM phase shift dependence on scan height

To understand the limits to the detection of small SPIONs with MFM, we experimentally analyzed the magnetic force sensitivity of MFM by examining the relationship between MFM phase shift and distance between the probe and the sample in hover mode imaging (Figure 3). To mimic various biological media we embedded SPIONs under a 30 nm layer of PS on mica. From Figure 3, the phase contrast in the MFM images is observed to increase with distance between the probe and the sample surface (scan height). Figure 4A shows the plot of MFM phase-shift (degrees) versus particle size (nm) for each scan height of $SiO_2$ SPIONs covered with PS as seen in Fig. 3. Best fit lines show positive linear trends for these data sets. This relationship is plotted in Figure 4B. The data shows that for this height range as scan height increases, MFM phase shift measured also increases reaching a plateau at approximately 300 nm.

We observe the increase of phase-shift signal with the increase of SPIONs size (Fig. 4A) and therefore magnetic moment. Interestingly we observe the increase of magnetic signal when the distance between the probe and the sample increases from 50 to 200 nm and then it follows the plateau at 250 and 300 nm (Fig. 4B). Therefore we show that MFM method has enough sensitivity to detect SPIONs at small as well as at larger separations between the probe and the sample.

## MFM of SPIONs in liquid

MFM imaging of $SiO_2$ coated magnetic particles in air has been reported before [15, 19]. In this work, for the first time we present MFM images of SPIONs in liquid environment. This step served as a proof of principle for using MFM in a liquid environment to image SPIONs in cells.

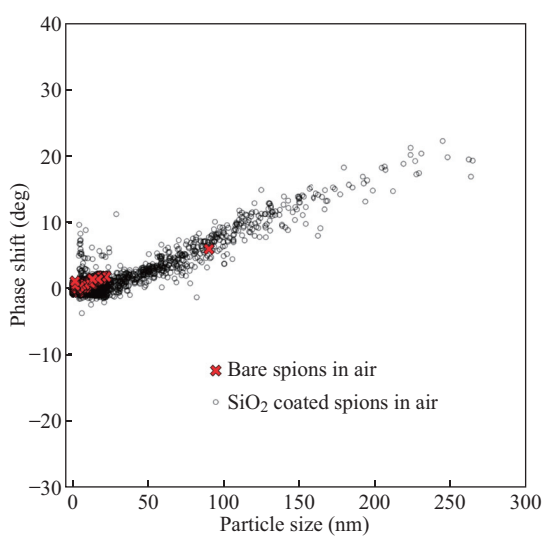

**Fig. 2** MFM phase-shift vs. particle size for bare and $SiO_2$ coated SPIONs. Phase-shift values obtained in MFM experiments on bare and $SiO_2$ coated SPIONs described in Figure 4.2. Phase shift (measured in degrees) versus SPION size (nanometers) is shown. A positive linear trend is observed for both bare and $SiO_2$ SPIONS. Phase shift values for bare SPIONs range from 0 to ~4 degrees and from 0 to ~20 degrees for $SiO_2$ coated SPIONs. MFM analysis for this data set was done in ambient conditions.

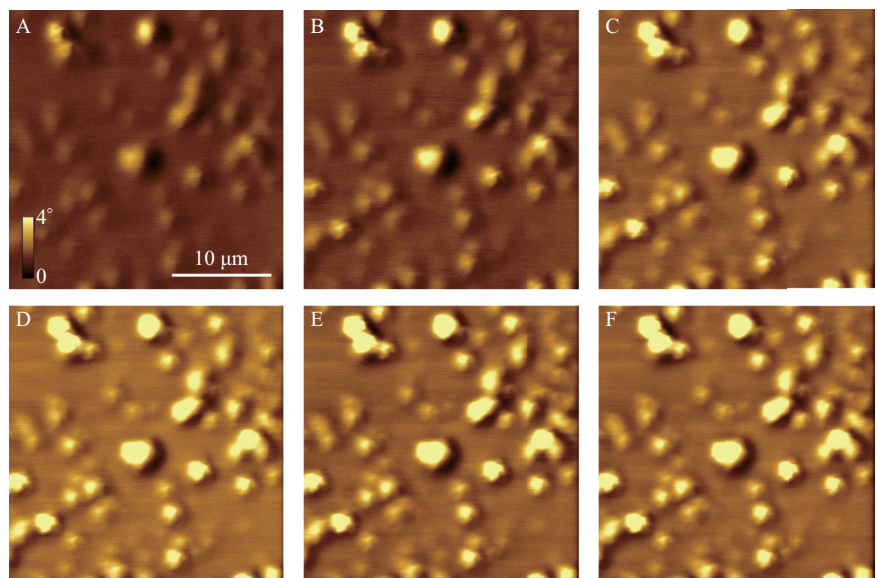

**Fig. 3** Consecutive MFM phase shift images taken at different scan heights for SiO$_2$ coated SPIONS covered with PS. Scan heights are (A) 50 nm; (B) 100 nm; (C) 150 nm: (D) 200 nm; (E) 250 nm; (F) 300 nm. Images were taken using a magnetic probe in the presence of an externally applied perpendicular magnetic field in ambient conditions. Colour-scale and scale bars shown in (A) apply to all images.

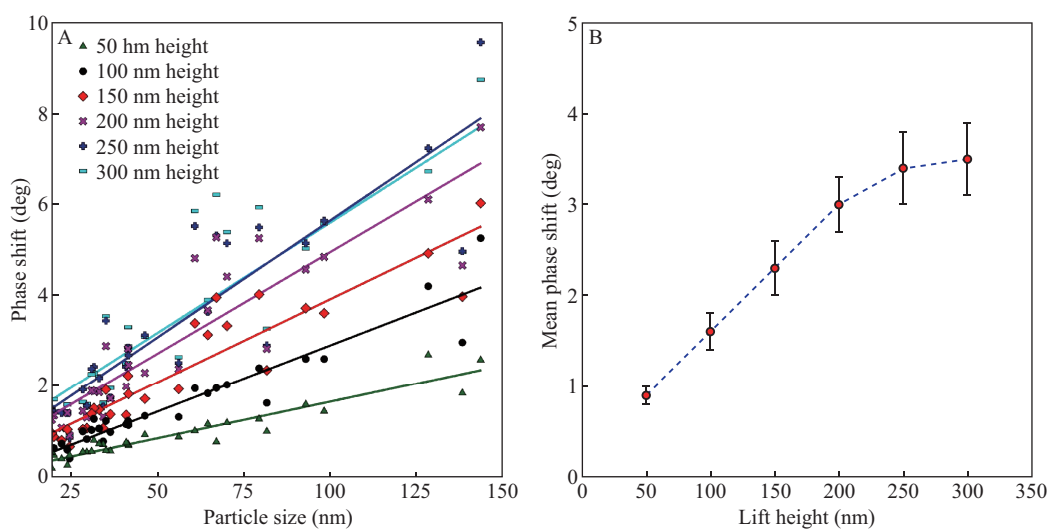

**Fig. 4** (A) MFM phase-shift (degrees) versus particle size (nm) is shown for each scan height of SiO$_2$ SPIONs covered with PS shown in Figure 3. Best fit lines show positive linear trends for these data sets. (B) Mean phase shift (for each set of SPIONs of different sizes) versus MFM scan height values for the phase images shown in Figure 3.

In this experiment SPIONs were coated with PMMA according to the protocol outlined in Section 2.2. Coating the sample with PMMA temporarily secured the nanoparticles to the substrate and prevented the SPIONS from being removed from the substrate when they were exposed to water.

Relatively small agglomerations of SPIONs covered with an approximately 30 nm layer of PMMA are shown in Figure 5. Considerably wider and more gradual structures are observed in when compared to the SPION structures seen in Figure 1. This comparison suggests that the SPIONs in Figure 5 are indeed coated with PMMA. In the MFM image, Figure 5 panel B, the same dipolar contrast that was observed in Figure 1 (panels B and D) can be seen. Fig. 5B shows clear contrast of SPIONs covered with PMMA and imaged in liquid. To the best of authors' knowledge, this is the first case of MFM imaging of SPIONs in liquid reported up to date.

## SPIONs on gold coated 3-MPTS functionalized mica

SPIONs are small nanoparticles; nevertheless we had problems depositing them on mica surfaces as they are easily washed away without special modification of the surface. Therefore we used chemical means of adhering the SPIONs to the surface of the substrate. The most promising method attempted to date for securing the SiO$_2$ coated SPIONs to the mica is the use of gold-coated mica substrates further modified with 3-mercaptopropyltrimethoxysilane (3-MPTS), an organosilane [20]. The key to this method is the formation of a covalent bond between the gold coated substrate and the SiO$_2$ coated SPION via the 3-MPTS molecule, specifically via its thiol functional group. Formation of this covalent bond will provide stability in air and liquid environments. An illustration of this deposition method as well as the chemical structure of 3-MPTS is shown in Figure 6.

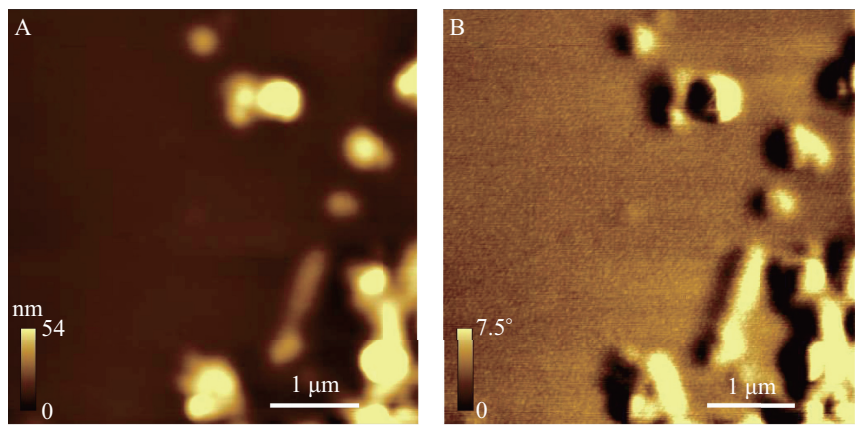

**Fig. 5** MFM in liquid image of SiO$_2$ coated SPIONs spin-coated with PMMA. Tapping-mode AFM topography image (A) and MFM phase image (B) of SiO$_2$ coated SPIONs spin-coated with PMMA taken at a lift height of 50 nm. Images were taken using a magnetic AFM probe in the presence of an externally applied perpendicular magnetic field in liquid (water). A spin motor with an applied voltage of 1V for 15 seconds was used to spin coat a 3% solution of PMMA (in toluene) in order cover the SPION sample with ~30 nm of PS. Colour-scale and scale bars are shown in the bottom of each panel.

This method is the most promising method attempted thus far for securing the SiO$_2$ coated SPIONs to the mica. In the MFM phase image shown in Figure 7B even better MFM contrast from the SPIONs can be observed, as compared to Figure 1D. Individual SPIONs can now be distinguished from within small aggregates with high level of magnetic detail. Therefore we conclude that mica substrates coated with gold serve as better substrates for SPION deposition, especially for AFM and MFM imaging.

The SPIONs in Figure 7 were observed even after the substrates were kept in aqueous solution for an hour followed by thorough rinsing. This demonstrates the formation of the covalent bond between the SiO$_2$ coated SPIONs and the 3-MPTS and its ability to secure these SPIONs to a gold-modified substrate in the presence of liquid. From Figure 7, it can be seen that successful SPION deposition only occurred for SPIONs in a water solvent. TMAOH has been shown to be a strong etchant of SiO$_2$ (~1 nm/min), which forms the coating around the SPIONs [5]. Since this etching process would have been occurring for the SPIONs in TMAOH, these particles were effectively uncoated bare SPIONs. As seen in Figure 6, this deposition process involving 3-MPTS is dependent upon the silanol functional groups present on the SiO$_2$ coating of the SPIONs. Thus, with the SiO$_2$ coating absent, deposition via 3-MPTS will not occur.

## Conclusions

Despite great progress and a rapidly advancing field of research there have always been problems associated with magnetic nanoparticles and their applications; overcoming immunological reactions, avoiding toxic responses to intravenously injected particles, proper clearance of particles, and the debate over whether or not to sacrifice more efficient uptake at the cost of negative side effects have been prominent issues within the field. Proper and relevant *in vitro*, and ultimately *in vivo*, characterization of the magnetic properties of these nanoparticles can now be added to this list.

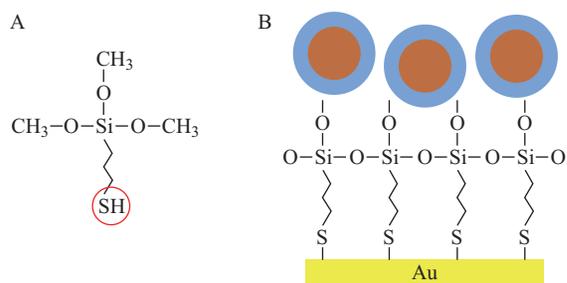

**Fig. 6** Deposition of SiO$_2$ coated SPIONs via (3-MPTS). (A) Chemical structure of 3-MPTS with the red circle indicating the thiol functional group. (B) Illustration of SiO$_2$ deposition with the blue region indicating the SiO$_2$ coating containing the silanol functional group.

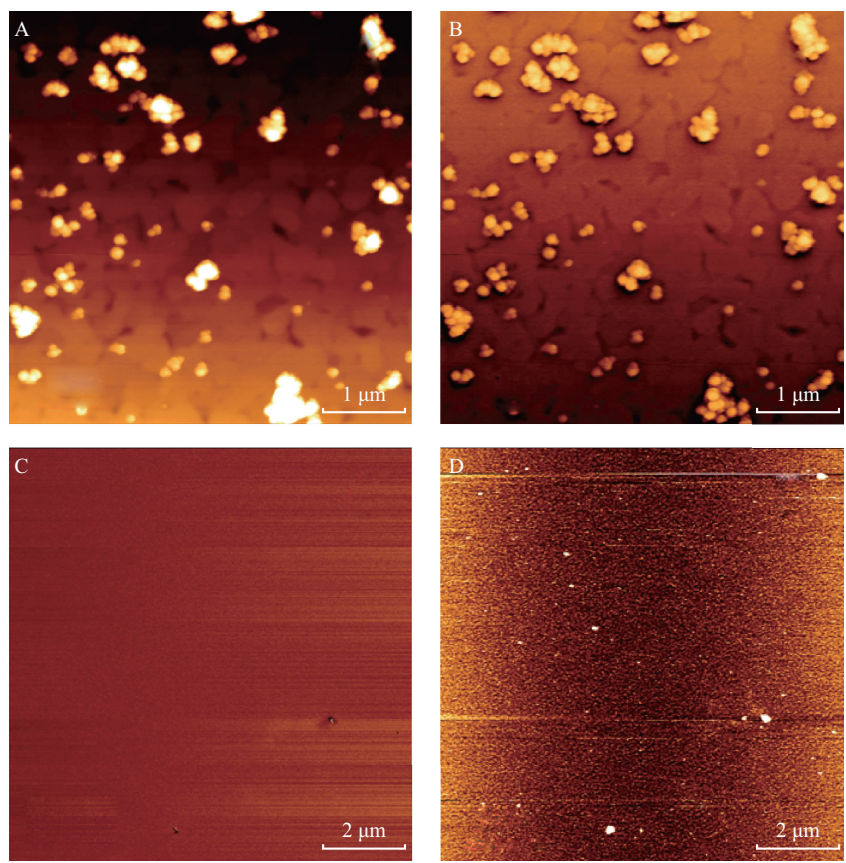

**Fig. 7** Comparson of SPIONs kept in water solvent (A, B) and TMAOH (C, D) on 3-MPTS functionalized gold coated mica. (A) and (C) are AFM topography images, (C) and (D) are MFM images. Images were taken using a magnetic AFM probe in the presence of an externally applied perpendicular magnetic field in ambient conditions. Scale bars are shown in the bottom left of each panel. Successful SPION deposition was only observed for the SPIONs kept in water solvent.

We demonstrated that MFM can be used to image SPIONs in air, in water and inside the polymer films. We report that surface modification with (3-mercaptopropyl)-trimethoxysilane significantly improves adsorption and distribution of SPIONs on gold surfaces. Our results show feasibility of using MFM for the detection of magnetic nanoparticles within cells without any labeling or modifications, thus demonstrating that MFM can be used as a potential magnetic analogue for fluorescence microscopy. These results may also add to further developments of SPIONs and their applications in biomedicine.

## Acknowledgments

We acknowledge Timothy Leshuk for assisting in synthesizing the SPION particles and the University of Waterloo Institute for Nanotechnology for help with sample preparation. This work was funded by Natural Science and Engineering Council of Canada (NSERC).

...